\documentclass[aps,floats,prd,psfig]{revtex4}
\input epsf
\usepackage{graphicx}% Include figure files

\newcommand{\beq}{\begin{equation}}
\newcommand{\beqa}{\begin{eqnarray}}
\newcommand{\eeq}{\end{equation}}
\newcommand{\eeqa}{\end{eqnarray}}

\newcommand{\lsim}{\lesssim}
\newcommand{\gsim}{\gtrsim}

\newcommand{\vect}[1]{\mbox{\boldmath${#1}$}}
\newcommand{\lmk}{\left(}
\newcommand{\rmk}{\right)}
\newcommand{\lnk}{\left\{ }

\newcommand{\rnk}{\right\} }
\newcommand{\lkk}{\left[}
\newcommand{\rkk}{\right]}
\newcommand{\lla}{\left\langle}

\newcommand{\rra}{\right\rangle}

\newcommand{\veN}{\vect N}

\newcommand{\bpsi}{\bar \psi}

\newcommand{\cq}{{\cal Q}}

\begin{document}
%\baselineskip 7mm
%\if0
%\draft
\title{Prospects of  LIGO for  constraining inclination of merging compact
binaries associated 
with three-dimensionally localized short-hard GRBs} 

\author{Naoki Seto}
%\email{nseto@uci.edu}
\affiliation{Department of Physics, and Astronomy, 4186 Frederick Reines
Hall, University of California, Irvine, CA 92697
}
%\fi
\begin{abstract}
We study prospects of a method to constrain the inclination of a
 coalescing 
 compact  binary  by detecting its gravitational waves associated with a
 three-dimensionally  localized  (direction  and  distance) short-hard
 gamma-ray burst.  We take advantage of   a synergy of these two
 observations,  and  our method can be   applied   even with a  single
 interferometer.  For a
 nearly face-on binary the inclination angle $I$ can be constrained in
 the range $1-SNR^{-1} \le \cos I\le 1$ ($SNR$: the signal to
 noise ratio of gravitational wave detection), provided that the error
 of the distance estimation is negligible. This method would   help us to study   properties of the
 short-hard bursts, including   potentially collimated jet-like
 structures as indicated   by recent observation.         
\end{abstract}
\maketitle
%\fi

\section{Introduction}
The gamma-ray bursts have been  known to be divided into two classes, the
long-soft bursts and the short-hard bursts (SHBs).  While the
former are 
likely to be produced at  explosions of massive stars in star forming
galaxies typically at high redshift $z\gsim 1$ \cite{Zhang:2003uk}, 
the nature of 
SHBs has been a long-standing mystery. However,  recent discoveries of
X-ray 
afterglows of SHBs by Swift and HETE satellites allowed us to localize
them accurately and rapidly enough  to specify their host galaxies and
finally determine their distances \cite{Gehrels,Villasenor:2005xj,Hjorth:2005kf,Fox:2005kv,Berger:2005rv}.

One of them, GRB 050724  was found in an elliptical galaxy at  $z=0.257$
with an old stellar population \cite{Hjorth:2005kf}, and   GRB 050509b is
likely to be in a 
similar galaxy at $z=0.225$ \cite{Berger:2005rv}. While GRB 050709 was
in a 
star forming galaxy at 
$z=0.160$,  its light curve excluded a supernova association
\cite{Fox:2005kv}.  
These results support that coalescing compact binaries (double neutron
stars (NS+NSs) or black hole-neutron star (BH+NS) systems) are the
promising 
origins of SHBs \cite{Eichler:1989ve}, though the estimated typical age of these binaries for
SHBs are longer than that of known NS+NSs in our galaxy \cite{Guetta:2005bb,Nakar:2005bs}.

Coalescing compact binaries are also promising sources of gravitational
radiation for LIGO and other ground-based interferometers
\cite{Thorne_K:1987}. A one year 
scientific run (S5) is ongoing with LIGO that has sensitivity to detect
NS+NSs to $\sim 15$Mpc \cite{ligo}. Recent theoretical analysis predicts
that the probability of a 
simultaneous  
detection of gravitational waves by LIGO and a SHB by Swift in one year
is $\sim 
30$\% for BH-NS merger and $\sim 10$\% for NS+NS (depending on the lower
end of the luminosity function of SHBs) \cite{Nakar:2005bs}. Therefore, we might soon
experience 
the first detection of  gravitational waves associated with a localized
SHB. 

The observed afterglows of two SHBs showed steeper power-law decays that
indicate SHBs have collimated jet-like structures \cite{Fox:2005kv,Berger:2005rv}, as found with long-soft
bursts.  The estimated beaming factor is $\sim 0.03$ for GRB 050709
\cite{Fox:2005kv} and
$\sim 0.01$ for GRB 050724 \cite{Berger:2005rv}. 
Here we defined the beaming factor as the fraction of $4\pi$
steradians into which  jets are emitted.
This enabled us to
estimate their  total
energies $\sim 3\times 10^{48}$erg  that is smaller than the
long-soft bursts by $\sim 2$ orders of magnitude. If a  SHB
is associated with a coalescing binary, it is likely that the
orientation of the jet is aligned with the angular momentum of the
binary that would be clearly imprinted on the observed gravitational
waveform. 

In this paper we propose a method with which the inclination of   a binary
will be interestingly constrained  as a synergy of three-dimensional
localization  by electro-magnetic waves (EMWs) and observation of
gravitational waves,  even using a single interferometer.  Therefore,
LIGO could provide us an important 
geometrical 
information to understand properties of SHBs.

This paper is organized as follows: in \S II we describe our method to
estimate inclination of binaries with using a single gravitational wave
interferometer. Expected error for our method is studied in \S
III. Then, in \S IV,
we extend our study for observation with  multiple detectors, such  as,
LIGO-VIRGO 
network. Brief discussions are presented in \S V.

\section{  GWs and constraint for  inclination}
For simplicity we use the restricted post-Newtonian description
\cite{Cutler:1994ys} with 
neglecting  precession  induced by spin that might be
important for 
BH-NS, but not for NS+NS \cite{Apostolatos:1994mx} (see also
\cite{Buonanno:2005pt} for recent analysis). The two polarization
waveforms in the principle 
polarization coordinate are given as \cite{Cutler:1994ys,Jara}
\beqa
h_+(t)&=&B f^{2/3} (1+V^2) \cos(\Phi(t)),\label{hp}\\
h_\times(t)&=&B f^{2/3} (2V)\sin(\Phi(t)), \label{hc}
\eeqa
where 
$\Phi(t)=2\pi\int_{t_I}^t f(t',M_c,...,t_I)dt'+\varphi_c$
 is the phase of the waves,
and $t_I$ and $\varphi_c$ are constants. The parameter $V$ is defined   by
\beq
V\equiv \cos I 
\eeq
 with the inclination angle $I$ shown in figure 1, and is the primary
target in this paper. The chirp mass $M_c$ is the most important
parameter to characterize the time evolution of the frequency
$f=1/2\pi ~d\Phi/dt$ and given by two (redshifted) masses of the binary as
$M_c=m_1^{3/5}m_2^{3/5}(m_1+m_2)^{-1/5}$.
The intrinsic amplitude $B$ at the quadruple order is given by the chirp mass
$M_c$ and the distance (more precisely, luminosity distance) $r$ to the
binary  \cite{Thorne_K:1987,Schutz:gp},
\beq
B=2\frac{G^{5/3}M_c^{5/3}\pi^{2/3}}{rc^4}.
\eeq
Response $h(t)$ of the interferometer to the two polarization modes is
written in terms of the beam-pattern functions  $F_+$ and $F_\times$ as
\beq
h(t)=F_+h_+(t)+F_\times h_\times (t).\label{h}
\eeq
The beam-pattern functions are determined by three angles $\theta,\phi$ and
$\psi $ as \cite{esta}
\beqa
F_+&=&a_1\cos2\psi-a_2\sin2\psi, \\
F_\times&=&a_1\sin2\psi+a_2\cos2\psi
\eeqa
with
$a_1\equiv \frac12(1+\cos^2\theta)\cos2\phi$ and  $a_2\equiv
\cos\theta\sin2\phi$. The angles $(\theta,\phi)$ represent the direction
of the binary in the polar coordinate attached to the interferometer as in
figure 1, and $\psi$ is the polarization angle that fixes the axial
direction of the vector $\vect L$ around the direction $\vect N$ (see
{\it e.g.} figure 1 in \cite{Apostolatos:1994mx}).

\begin{figure}
  \begin{center}
\epsfxsize=10.cm
\begin{minipage}{\epsfxsize} \epsffile{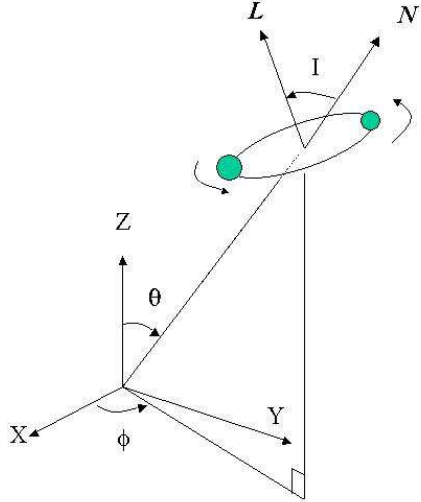} \end{minipage}
 \end{center}
  \caption{Configuration of a binary  and an interferometer ($X$ and
 $Y$-axes; 
 directions of its two arms). The unit vector $\vect N$ represents the
 direction of the binary, and the unit vector  $\vect L$ is the orientation of its angular
 momentum. The angle $I$ is the inclination with $\cos I={\vect L}\cdot {\vect N}\equiv V$.  The polarization angle $\psi$ specifies  the  direction of $\vect L$
 around the vector $\vect N$. }
\end{figure}

%With eqs.(\ref{hp})(\ref{hc}) and (\ref{h}) the observed amplitude of
%the detector's signal is written as
When we take Fourier transformation of the response $h(t)$,  the result
is formally given as
\beq
{\hat h}(f)=AR(V,\psi,\theta,\phi)f^{-7/6} \exp[i \Psi(f)].
\eeq
Here we used the stationary phase approximation, and the phase $\Psi(f)$
is a real function.
The overall
amplitude of the signal 
$A R(V,\psi,\theta,\phi)$ is given by  \cite{Jara}
\beq
R=\sqrt{\lnk(V^2+1)  F_+\rnk^2+(2VF_\times)^2}, \label{re}
\eeq
and 
\beq
A=\sqrt{\frac{5}{96}}\frac{G^{5/6}}{\pi^{2/3} c^{3/2}}\frac{M_c^{5/6}}{r}.
\eeq
The quantity $R$ is a complicated function of four angular variables
$(I,\psi,\theta,\phi)$, and we cannot solve them separately only with a
single interferometer. Even if the direction $(\theta,\phi)$ is known
({\it e.g.} from EMW observation), we need at least two not-aligned
interferometers to solve $V$, $\psi$ and $r$ from  observed amplitudes
and  the relative phases (see \cite{Kobayashi:2002md} and references
therein). 

\begin{table}

  \begin{center}
    \caption{Determination of $Q\equiv (A R)/A R_{max}$ }
    \begin{minipage}{9.0cm}
      \begin{tabular}{c|c|cc} \hline\hline
         notation & observation & error \\ \hline
        
        $A$ & $M_c$: GW (chirp) & $\sim 0$& \\
        
                 $(\propto M_c^{5/6}/r)$     & $r$: EMW (localization) & 
       $\Delta H_0/H_0 $, velocity&\\\hline

        $(AR)$ &  GW (amplitude) & $\sim (SNR)^{-1}$& \\\hline
 
        $R_{max}$ & $(\theta,\phi)$: EMW (localization) & $\sim0 $& \\\hline
        
        \hline
      \end{tabular}
      \vspace*{-0.4cm}
    \end{minipage}
  \end{center}
\end{table}

Now we assume a situation that the three dimensional position
$(r,\theta,\phi)$ of a coalescing binary is determined through the
afterglow of SHB associated 
with coalescing (ispiral) gravitational waves detected by LIGO. We start
with the case for a  single  interferometer.  With
gravitational  
wave observation the chirp mass $M_c$ can be determined well by  the 
frequency evolution \cite{Schutz:gp}. Thus we can estimate the intrinsic
amplitude 
$A\propto M_c^{5/6}/r$ by combining the chirp mass and the distance $r$
from EMW localization. From 
gravitational wave data we can also get the combination $(AR)$ as the
observed amplitude. We describe the method to constrain the parameter $V$
by dealing with these observed quantities. Basic aspects are summarized
in table 1.
Firstly, we calculate the maximum value of $R(V,\psi,\theta,\phi)$ for
the observed direction $(\theta,\phi)$ of SHB. This is realized with
the 
face-on configuration with $|V|=1$.  In this case the
polarization angle $\psi$ does not have meaning due to the symmetry of
the geometry, and $R_{max}$ is independent on $\psi$, 
\beq
R_{max}\equiv R(1,\psi,\theta,\phi).
\eeq
Secondly, we take the ratio $Q$ of the observed gravitational wave
amplitude 
$(AR)$ to the estimated combination $A\cdot R_{max}$ as
\beq
Q=\frac{(AR)}{A\cdot R_{max}}=\frac{R}{R_{max}}.\label{defq}
\eeq
The ratio $Q$ is written by
\beq
Q=\sqrt{\frac14 (1+V^2)^2 \cos^2(2
\psi+\gamma)+V^2
\sin^2 (2\psi+\gamma)}. \label{q}
\eeq
with $\gamma=\arctan(a_2/a_1)$.
This expression is valid also in the limit $(\theta,\phi)\to (\pi/2,
\pi/4)$  
where both the denominator and the numerator in  eq.(\ref{defq}) vanish,
as the interferometer becomes insensitive to the incident gravitational waves.
In figure 2 we show the profile of $Q$ with $\theta=\phi=0$. The
ratio $Q$ takes a same value for $\pm V$ and, is periodic along the
$\psi$-direction with period $\pi/2$. For a different combination
$(\theta,\phi)$ the profile $Q$ in figure 2  shifts to  the
$\psi$-direction  as characterized by the angle $\gamma/2$
in eq.(\ref{q}). Therefore, without loss of generality, we
can study the constraint on $V$ for arbitrary $(\theta,\phi)$ only
using figure 2 with which the constrained region for $V$ can be easily
read with a given  $Q$. Note that the modes higher than the
quadrupole order might slightly change this contour map and also
affect the estimation of the quadrupole amplitude $A$. These
aspects are not peculiar to our method, but common to data  analyses for
gravitational wave astronomy for coalescing binaries.

\begin{figure}
  \begin{center}
\epsfxsize=11.cm
\begin{minipage}{\epsfxsize} \epsffile{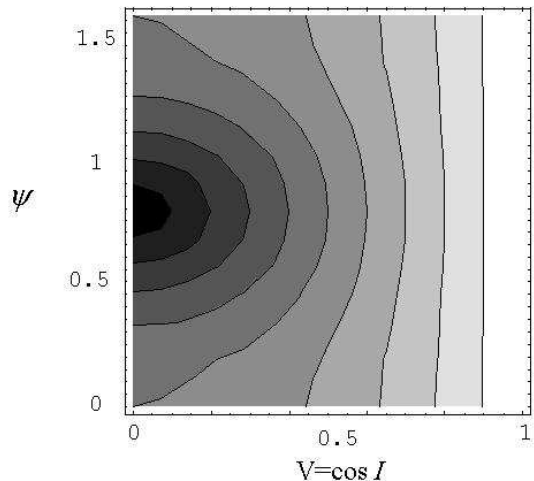} \end{minipage}
 \end{center}
  \caption{ The contour plot for $Q$ as a function of $V$ and $\psi$
 with $\theta=0$ and 
 $\phi=0$. Contour levels 
 correspond to 0.1 (black) to 0.9 (white). $Q$ is periodic along
 $\psi$-direction with period $\pi/2$, and $Q=1$ for $V=1$ (face-on
 configuration). The contour plot $Q$ shifts to $\psi$-direction for  different combinations 
 $(\theta,\phi)$.}
\end{figure}

For a given value of $Q$, the maximum of $V$ is realized with the choice
$\psi=0$ in figure 2, and the minimum is with $\psi=\pi/4$. 
In figure 3 (solid curves) we show the constrained region for $V$ as a
function of $Q$. The upper curve is $V=Q$ and the lower one is
$V=\sqrt{2Q-1}$.  Thus the parameter $V$ is constrained by 
$\sqrt{\max(2Q-1,0)}\le V \le Q$.
For the ratio $Q$ close to 1, the binary must be almost face-on, as this
is the 
only configuration to realize the maximum amplitude  $Q=1$. In
contrast, for $Q\le 1/2$, the parameter  $V$ can take the range $[0,Q]$.
The allowed region for $V$ is very small for $Q\sim 1$, when the binary
is nearly face-on and SHB is expected to be luminous with  collimated
jet-like structures. Even with a smaller $Q$ ({\it e.g.} $Q=0.6$) we can
correctly discriminate that the SHB is completely off-axis.

So far we have studied case with single interferometer. By using three
widely separated interferometers the direction $(\theta,\phi)$ of a
binary can be determined  from the time delays of the
gravitational wave signals, and the angles $(I,\psi)$ and the distance
$r$ can be also estimated only from gravitational wave observation
\cite{Cutler:1994ys}. As  commented earlier,  we can, in principle,
solve two orientation angles ($I,\psi$) and the distance $r$ by using two
not-aligned interferometers for a binary with known direction. We will
see this in \S IV.

  LIGO has essentially two
interferometers. One is 
at Hanford, 
Washington, and another is at Livingston, Louisiana. While the 
separation between them is  $\sim 3000$km (corresponding to $\sim
27^\circ$ measured from 
the center of the earth), they are nearly parallel to increase 
correlation of burst signals \cite{Cutler:1994ys}.
As a result, it is not easy to observe two polarization modes separately
and thereby estimate the parameter $V=\cos I$ well with using two LIGO
interferometers, compared with using combination such as LIGO-VIRGO
network. In \S IV we will return to analyses with multiple
interferometers. But our studies for a single interferometer would
provide rough outlook for determination of the inclination with nearly
aligned interferometers as LIGO.
\if
  From two observed values $Q_H$ and $Q_L$ we
can solve $V$ and $\psi$ separately. Here, the polarization angle
$\psi\equiv \psi_H$ is defined with the Hanford coordinate.  Compared
with $\psi$, the polarization angle $\psi_L$ with the   Livingston
coordinate has a constant off-set that is determined by the direction of
the source $\vect N$. In addition, the apparent directional angles $(\theta,\phi)$ are different in these two coordinates. If we plot
$Q_H$ and $Q_L$ as functions of $(V,\psi=\psi_H)$ for a given source
direction $\vect N$,
the profiles are again obtained by constant shifts of figure 2 along the
$\psi$-direction.  The intersection of contour lines for two observed
values 
$Q_H$ and $Q_L$ in $(V,\psi)$-plane is the solution for the orientation of
the binary.  In addition to two
amplitudes, the phase difference between two interferometers  has
 information of the orientation, while we do not study it in this paper.
\fi
\begin{figure}
  \begin{center}
\epsfxsize=11.cm
\begin{minipage}{\epsfxsize} \epsffile{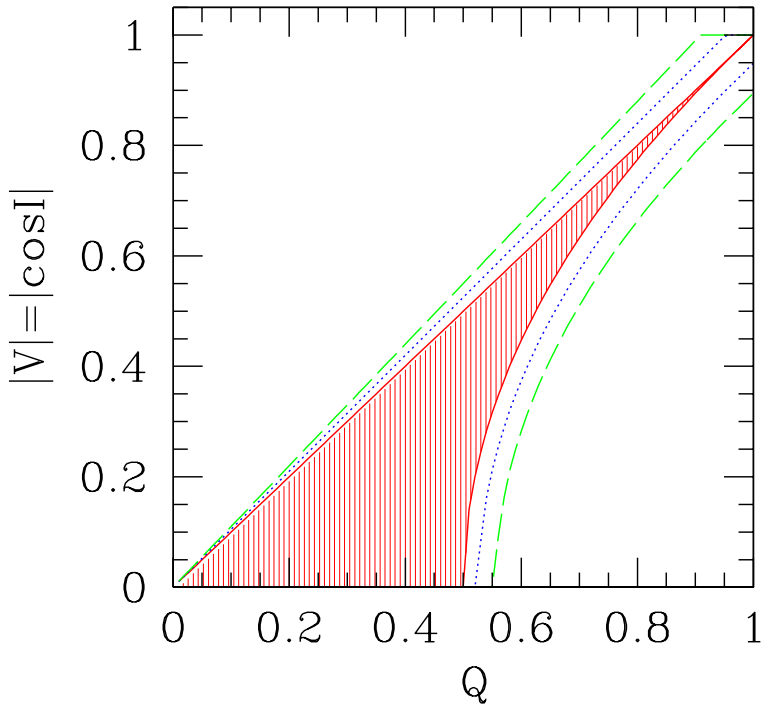} \end{minipage}
 \end{center}
  \caption{ Constraint for $V=\cos I$ with  given ratio
 $Q$. Without error  $\Delta Q=0$, the parameter $V$
 must 
 be within the shaded region determined by two solid curves $V=Q$ and
 $V={\sqrt{\max(2Q-1,0)}}$ . These two curves are obtained by cutting
 the surface in figure 2 along 
 $\psi =0$ and $\psi=\pi/4$. The dotted curves show the allowed
 region with 5\% error for $Q$ ($\Delta Q/Q=0.05$), and the dashed curves
 show the region with 10\% error  ($\Delta Q/Q=0.10$). With a nearly
 face-on 
 configuration we have $\Delta V\in [1-\Delta Q, 1]$. }
\end{figure}

\section{   Observational error}
In actual observation, we cannot determine the ratio $Q$ without error
$\Delta Q$. Here we analyze its effects  for  constraining  
$V$.  
As we discussed, the ratio is obtained by three observed values $A$,
$R_{max}$ and $(AR)$ (see table 1). We can formally write down the
relative error for $Q$ as
\beq
\frac{\Delta Q}{Q}=\frac{\Delta (A R)}{(A R)}+\frac{\Delta
A}{A}+\frac{\Delta R_{max}}{R_{max}}.\label{base}
\eeq
We are dealing with a situation when the afterglow of SHB is observed
and the direction $(\theta,\phi)$ is determined very  well ({\it e.g.}
$\lsim 1$arcsec level). Therefore, the term ${\Delta R_{max}}/{R_{max}}$
is negligible compared with other two terms.

If the orbital precession of the binary is not significant, the
estimation error
for the  observed gravitational wave  amplitude $(AR)$ has little
correlation with errors for other parameters 
related to the phase, such as, the chirp mass. Thus we have
\beq
\frac{\Delta (A R)}{(A R)}\simeq(SNR)^{-1} \label{amp}
\eeq
with $SNR$ being the signal to noise ratio for the detected
gravitational wave signal \cite{Cutler:1994ys}.
The typical detection threshold for a coalescing binary is
$\sim 8$. If we use the temporal information from the observed SHB, the
threshold somewhat decreases (but less then a factor of  2)
\cite{Kochanek:1993mw}. 

For the intrinsic amplitude $A$ we need the distance $r$ from EMW
observation and the chirp mass $M_c$ from gravitational wave data. The
latter can be determined very accurately, as it is the primary parameter
for the time evolution of the gravitational wave phase whose information
is crucial for  detecting gravitational waves with  matched filtering
method 
\cite{Thorne_K:1987,Schutz:gp}. In this paper we neglect  estimation
error for chirp mass. Under this prescription our results do not depend
on the order of the post-Newtonian expansion in the restricted
post-Newtonian approach without orbital precession.  For a
very 
nearby SHB we might estimate its distance $r$ with various astronomical
data.  If we use the redshift-distance relation (Hubble law; $r= 
cz /H_0$ at low redshift) to convert its observed redshift $z$ to the
distance $r$,  
 the estimated distance $r$ might be significantly contaminated by the
peculiar velocity of the host galaxy. At distance $r\sim 300$Mpc (NS+NSs
detectable with LIGOII) the uncertainty of the Hubble parameter $H_0$
could 
be a problem, but those of other cosmological parameters ({\it e.g.} the density
parameter $\Omega_0$) would not be important at these distances. The
recently reported value 
$H_0=71^{+4}_{-3}$km/sec/Mpc by WMAP team \cite{Spergel:2003cb} contains
$\sim 5\%$ error. As the 
distance error would be the dominate source of the error $\Delta A$,  we
have  
\beq
\frac{\Delta
A}{A}\simeq \frac{\Delta r}{r}.\label{dis}
\eeq
The error for the estimation of the gravitational wave amplitude $(AR)$
and the distance $r$ from EMW observation
would be independent. Then, using eqs.(\ref{base})(\ref{amp})(\ref{dis})
and putting $\Delta R_{max}/R_{max}=0$,  the typical value for  relative
rms 
error of $Q$ is given as  
\beq
\frac{\Delta Q_{rms}}{Q}\simeq\sqrt{\lmk \frac1{SNR}\rmk^2+\lmk\frac{\Delta
r}{r} \rmk^2} ~~~~({\rm typical ~case}).
\eeq 

Next we study how the estimation error $\Delta Q$ changes the
allowed region for  $V$. 
For a given $Q$ we expand its possible value in the band 
 $[\max(0,Q-\Delta Q),\min(1,Q+\Delta Q)]$ and solve corresponding $V$ in
$(V,\psi)$-plane (figure 2). 
Then we obtain the following result
\beq
\sqrt{\max\{2(Q-\Delta Q)-1,0\} }\le V \le \min(1, Q+\Delta Q).
\eeq
In figure 3 we added the allowed region for $V$ from the observed value
$Q$ whose relative error is $\Delta Q/Q =0.05$ (dotted curves) and
0.10 (dashed curves).
For a almost face-on binary ($1-V\ll 1$), we have $V\in [1-\Delta Q,
1]$. With signal to noise ratio $SNR\sim 10$ and the distance
error 
less than 5\%, the constraint becomes $V\in [0.87, 1]$.

\section{multiple detectors}

\if0
\subsection{extension of previous results}

When we use multiple detectors $(i=1,\cdots,N_d)$,  gravitational wave
amplitude 
$(AR_i)$ at each detector takes different values depending on the polarization angle
$\psi_i$ measured at each detector. If detector noises are not
correlated, the total signal to noise  ratio $SNR_{tot}$ is obtained by
summing individual ones  as follows
\beq
SNR^2_{tot}=\sum_{i=1}^{N_d} SNR_i^2,
\eeq
where each contribution $SNR_i$ is given by the amplitude $AR_i$ divided
by the noise level $n_i$ of detector $i$, as $SNR_i\propto
AR_i/n_i$. From the total signal to noise 
ratio $SNR_{tot}$, we can estimate an averaged amplitude $AR_{av}$
defined by 
\beq
(AR_{av})^2=\sum_{i=1}^{N_d} (AR_i)^2 w_i^2
\eeq 
with the weight factors $w_i\propto 1/n_i$ determined by the relative noise level and
$\sum_{i=1}^{N_d}w_i^2=1$.   Using
relations $\Delta (AR_i)/(AR_i)=(SNR_i)^{-1}$ for each detector, we can derive  $\Delta
(AR_{av})/(AR_{av})=(SNR_{tot})^{-1}$ for the estimation error $\Delta
(AR_{av})$ for the averaged amplitude with  multiple detectors.
Therefore, our results in the previous sections can be used for the cases
with 
multiple 
detectors by simply replacing $R\to R_{av}$ and $SNR \to SNR_{tot}$.
This argument to determine 
the inclination $V$ uses only the averaged amplitude $AR_{av}$  without
separating $V$ from  
the polarization direction $\psi$.
%But we can individually estimate them by measuring two polarization
%modes with un-aligned multiple
%detectors, as commented earlier. 

\fi

\subsection{Solving degeneracy}

In this subsection we discuss observational analysis to estimate the
inclination $V$ separately from the polarization direction $\psi$ with using
networks of detectors.  
Following Ref.\cite{Cutler:1994ys} we introduce two functions
$\sigma(\veN)$ and 
$\epsilon(\veN)$ that characterize sensitivity of  a network  to 
gravitational waves with various directions and polarizations. As
concrete examples of networks, we consider the following three
combinations; (i) 
LIGO-Hanford (4km), LIGO-Livingston and VIRGO, (ii) LIGO-Hanford (4km) and VIRGO,
(iii) LIGO-Hanford (4km) and  LIGO-Livingston. The total numbers of
detectors $N_d$ are three for case (i) and  two for cases (ii) and (iii).
For simplicity we assume that all   detectors have identical
sensitivity. The total signal to noise ratio for a binary with
directional  vectors
$(\veN,{\vect L})$, distance $r$ and a fixed chirp mass is written as
\beq
SNR_{tot}^2=\frac{r_0^2}{r^2} \sigma({\veN})\lkk
c_0(V)+\epsilon(\veN)c_1(V)\cos(4\bpsi)\rkk, \label{snr}
\eeq
where $\bpsi$ is the polarization angle of the binary angular momentum
${\vect L}$ measured from  a preferred direction that is determined by
configuration of 
each network and also fixes two orthogonal polarization bases in the present analysis. The combination $r_0/r$ represents the signal to noise
ratio of single detector for a face-on ($|V|=1$) binary at  perpendicular
direction to  the detector plane ($\theta=0$ in figure 1).
For a binary neutron stars with chirp mass $1.2M_\odot$ the
distance $r_0$ corresponds to $\sim$5000Mpc for advanced LIGO. Here we
used 
numerical 
results given in \cite{Dalal:2006qt} based on the noise curve for
advanced LIGO with its  wide band setting \cite{adv}. Using relative
sensitivities of LIGO and advanced LIGO to double neutron stars ({\it
e.g.} \cite{Cutler:2002me}), the distance $r_0$ for LIGO is $\sim
320$Mpc. Note that these distances are for $SNR=1$ with an optimal
configuration of a binary and  a single interferometer.

The function $\sigma(\veN)$ shows sensitivity of a network to
gravitational wave from  direction $\veN$ with averaged polarization,
and generally takes values in $0\le \sigma(\veN) \le N_d/2$ ($N_d$:
number of detectors with identical sensitivity). For our
three networks its maximum value is 1.04 for case (i), 0.65 for case
(ii) and 0.94 for case (iii), while its minimum value is 0.148, 0.118 and
0.0040 respectively. As two LIGO detectors are nearly aligned, they are
not effectively complimentary to decrease the 
blind directions with $\sigma(\veN)\ll 1$. 

The function $\epsilon(\veN)$ ($0\le \epsilon \le 1$) represents
asymmetry  of a  network with respect to its sensitivity to   two
orthogonal 
polarization modes coming from direction $\veN$. For a direction
$\veN$ with 
$\epsilon (\veN)
\sim 1$ a network is sensitive only to one polarization mode, while two modes are
measured with a similar sensitivity for $\epsilon(\veN)\sim 0$. With a single
detector we identically have $\epsilon(\veN)=1$. For our three networks,
this parameter can take the maximum value $\epsilon(\veN)=1$ for some
directions.   

In eq.(\ref{snr}) two functions $c_0(V)$ and $c_1(V)$ are defined by
\beq
c_0(V)=\frac{(1+V^2)^2}4+V^2, ~~~~c_1(V)=\frac{(1+V^2)^2}4-V^2. 
\eeq
For a face-on
binary with $|V|=1$, we have $c_1=0$ and the expression (\ref{snr}) does not
depend on the polarization angle $\bpsi$, as expected.

Next we evaluate magnitudes of estimation errors $\Delta \alpha_i$ for
fitting parameters $\alpha_i=(r,V,\bpsi)$ that are related to amplitude
of the 
gravitational waveform. We use  the Fisher matrix approach that
basically uses linear responses of the waveform to variations of fitting
parameters.  In other words,  the first derivatives of the
waveform by the fitting
parameters are used to evaluate the expected errors.  
Under this approach,  correlation between the parameters
$\alpha_i$ and those related to frequency evolution are  practically
negligible in the restricted post-Newtonian approach \cite{Cutler:1994ys}.

With using eqs.(4.38) and (4.39) in Ref.\cite{Cutler:1994ys} and integrating
out the information of the phase constant $\varphi_c$ (see arguments
following eq.(2)),  the covariance of estimation errors
for parameters $(r,V,\bpsi)$ are given as
\beqa
\lla \Delta \bpsi \Delta \bpsi \rra &=&E \lkk
1+6V^2+V^4+(1-V^2)^2\epsilon({\veN})\cos(4\bpsi) \rkk/(1-V^2)^2\label{v1}\\
\lla \Delta \bpsi \Delta r \rra &=&-2E r \epsilon({\veN})\sin(4\bpsi)\\
\lla \Delta \bpsi \Delta V \rra &=&0\\
\lla \Delta r \Delta r\rra &=&4Er^2
(1+V^2-(1-V^2)\epsilon({\veN})\cos(4\bpsi)) \label{dis}\\ 
\lla \Delta r \Delta V\rra &=&2ErV
(3+V^2-(1-V^2)\epsilon({\veN})\cos(4\bpsi))\\ 
\lla \Delta V\Delta V\rra  &=&E
\lkk 1+6V^2+V^4+\epsilon({\veN})\cos(4\bpsi)(1-V^2)^2\label{n} \rkk
\eeqa
with
\beq
E=\frac{r^2}{r_o^2 \sigma (\veN) (1-\epsilon(\veN)^2)(1-V^2)^2}.\label{E}
\eeq
Equation (\ref{dis}) for the distance error is same as eq.(4.41) in
Ref.\cite{Cutler:1994ys}.  In the next subsection we will compare the
estimation errors 
for the 
inclination under various situations, and hereafter, we denote
the expression (\ref{n}) as $\lla \Delta V\Delta V\rra_N$  for
notational clarity. The errors 
become  large for binaries close to face-on through 
the denominator of the factor $E$. This is because the derivatives of the gravitational
waveform with  respect to the parameters $r$ and $V$ become linearly
dependent at 
$|V|=1$, and they are degenerated at parameter fitting
\cite{Markovic:1993cr}. In the same 
manner the error becomes large for sky
 directions with $\epsilon(\veN)\sim
1$, as 
we can measure only one polarization mode that are not enough to
determine parameters $(r,V,\bpsi)$ separately. However, without
depending on the 
values $\epsilon(\veN)$ and $V$, the combination
$SNR_{tot}\propto 
[(c_0(V)+\epsilon(\veN)c_1(\veN)\cos(4\bpsi))]^{1/2}/r$ can be
 determined well
and its error is given  only by
the the total signal to noise ratio as $\Delta
(SNR_{tot})/(SNR_{tot})=(SNR_{tot} )^{-1}$.
The situation is same as in the previous
sections for single  detector. We can also confirm  this  with
directly using equations from (\ref{v1}) to
(\ref{E}). 

Now we study the case that the distance $r$ to a binary is given from
EMW observation, assuming its error is negligible. After straightforward
calculation with eqs. (\ref{v1}) to
(\ref{E}), the estimation errors for parameters $(\bpsi,V)$ are given
by a 
$2\times 2$ 
Fisher matrix, and the  $V-V$ component becomes
\beq
\lla \Delta V \Delta V\rra_{r}=\frac{r^2 (1-\epsilon({\veN})^2\cos^2(4\bpsi))}{r_0^2
\sigma({\veN}) (1-\epsilon({\veN})^2)(1+V^2-\cos(4 \bpsi)\epsilon(\veN)(1-V^2))}.\label{r}
\eeq
Here the suffix $r$ represents  the condition that the distance $r$ of
the binary is
given. 
Note that the singular behavior at $|V|=1$ disappeared. This is because  we do not
need to solve $r$ and $V$ simultaneously only from gravitational wave
observation. 
However, the error $\lla \Delta V \Delta V\rra_{r}^{1/2}$ for the
inclination still becomes
large with $\epsilon \to 1$. 
To deal with the situation  we introduce a new parameter $\cq$ defined by
\beq
{\cal Q}\equiv \frac{AR_{tot}}{AR_{max}}=\lmk \frac{c_0(V)+\epsilon(\veN) c_1(V)
\cos(4\bpsi)}{ c_0(1)+\epsilon(\veN) c_1(1)
\cos(4\bpsi)}\rmk^{1/2}=\lmk \frac{c_0(V)+\epsilon(\veN) c_1(V)
\cos(4\bpsi)}2 \rmk^{1/2},
\eeq
and change variables from $(\bpsi,V)$ to
$(\cq,V)$.
The parameter $\cq$ for multiple detectors is a simple generalization of
the 
previous  
parameter $Q$ in eq.(12) defined for single detector.
In figure 4 we show the allowed region of the combination $(\cq,V)$ for
given network parameter $\epsilon$.
 For a direction with
$\epsilon=1$ the allowed region coincides
with that for $Q$-$V$ given by the shaded 
region in figure 3. When we decrease $\epsilon$
form 1 to 0,   the region shrinks as in figure 4,
%For given $V$, the parameter $\cq$
%can move in the range bounded by 
%$[(c_0(V)-\epsilon c_1(V))/2]^{1/2}$ and $[(c_0(V)+\epsilon
%c_1(V))/2]^{1/2}$. 
and it
becomes 
a single curve $\cq=(c_0(V)/2)^{1/2}$ at $\epsilon=0$.

Next we  consider a Gaussian probability
distribution 
function (PDF) 
$P(\cq_x,V_x|\cq,V)$ of the estimated parameters $(\cq_x,V_x)$
around their true values $(\cq,V)$.  We calculate  the covariances of
errors for parameters $(\cq,V)$ by using
 the $2\times2$ Fisher
matrix for the original  combination $(\bpsi,V)$.
Even if the error $\lla \Delta V \Delta V\rra_{r}^{1/2}$ is large at
$\epsilon (\veN)\sim 1$ as shown in eq.(\ref{r}), the parameter $\cq$
can
be estimated relatively
well, as in the case for single detector.
For example, with a $1-\sigma$ error ellipsoid that is stretched toward
$V$-direction  as shown in figure 4, we
can estimate the inclination $V$ much better by limiting the combination
$( \cq_x,V_x)$
into its allowed region discussed above. This difference shows one of 
potential problems of a simple evaluation based on Fisher matrix without
dealing with global domain of the fitting parameters. 
Therefore, we define a new PDF $P_C(\cq_x,V_x|\cq,V)$ with the following
two steps.  We firstly
narrow the
parameter space $(\cq_x,V_x)$ of the original Gaussian distribution
 $P(\cq_x,V_x|\cq,V)$  using the $\cq$-$V$ constraint in figure 4, and secondly
renormalize the PDF appropriately.   Then we calculate the estimation
error 
$\lla 
\Delta V \Delta V \rra_C$  as
\beq
\lla \Delta V \Delta V \rra_C=\int dV_x d\cq_x P_C(\cq_x,V_x|\cq,V) (V_x-V)^2.
\eeq 
Here the suffix $C$ represents to use
the constraint.
\subsection{Monte Carlo analysis}

\begin{figure}
  \begin{center}
\epsfxsize=11.cm
\begin{minipage}{\epsfxsize} \epsffile{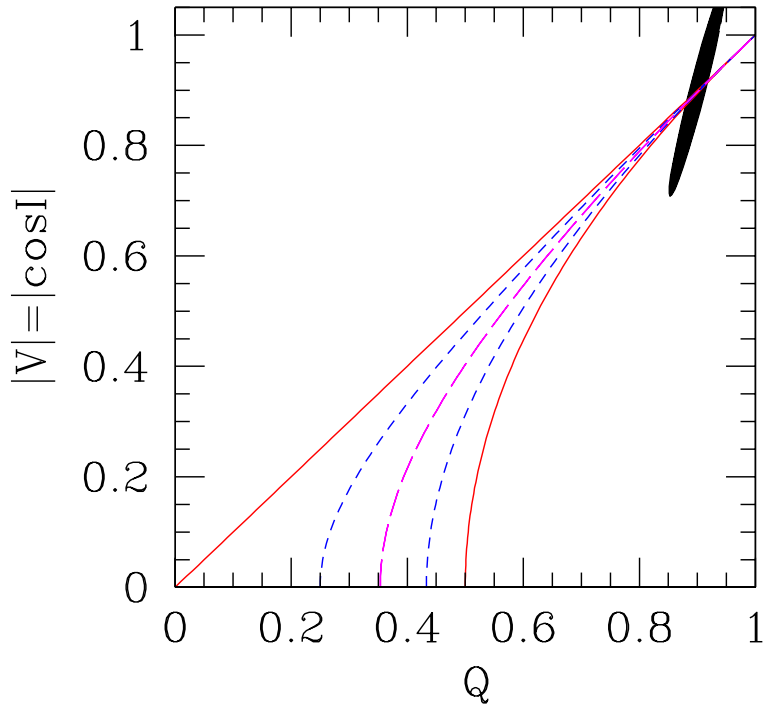} \end{minipage}
 \end{center}
  \caption{  Constraints for $V$ and  $\cq$ for given parameter
 $\epsilon(\veN)$ that represents asymmetry of the sensitivity of a
 network to 
 two polarization modes coming from  direction $\veN$. For $\epsilon=1$ the combination  $(\cq,V)$ is  in the region bounded by two
 solid curves.   This region is same as the shaded
 region in figure 3 for single detector. The short-dashed curves are for
 the boundary of the allowed region $\epsilon=0.5$.  For $\epsilon=0$ the allowed region is on the long-dashed line
 $\cq=[c_0(V)/2]^{1/2}$. The long ellipsoid around
 $(\cq,V)=(0.9,0.9)$ is an example of $1-\sigma$ error ellipsoid
 described in the main text.}  
\end{figure}

\begin{figure}
  \begin{center}
\epsfxsize=10.cm
\begin{minipage}{\epsfxsize} \epsffile{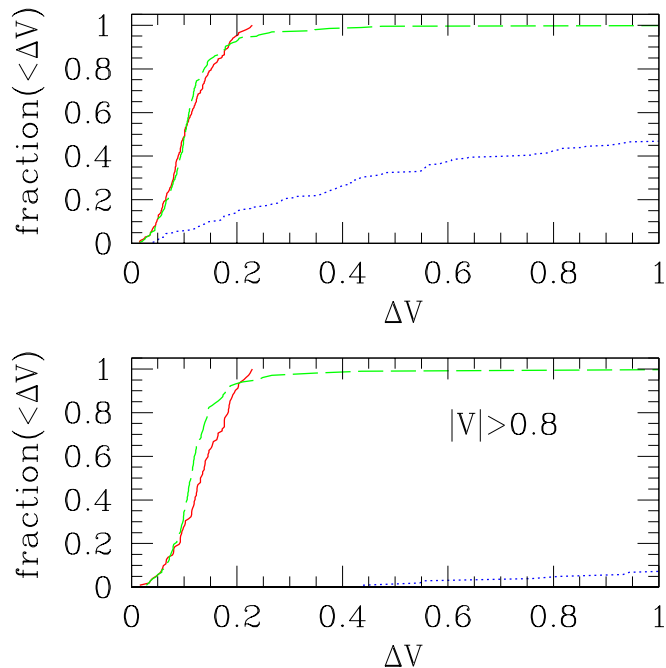} \end{minipage}
 \end{center}
  \caption{ cumulative distribution of the estimation errors $\Delta V$ for
  binaries 
 with using two LIGO  and  VIRGO detectors (case (i)). The solid
 curves are results for  $\lla \Delta V \Delta V\rra_C^{1/2}$ (known
 distance from EMW observation and $\cq$-$V$
 constraint). The 
 dashed curves are given by the 
 Fisher matrix approach  $\lla \Delta V \Delta V\rra_r^{1/2}$ (known
 distance  from EMW observation). The dotted curves are
 for $\lla \Delta V \Delta V\rra_N^{1/2}$ (unknown distance from EMW observation). The
 lower panel is for binaries with $|V|>0.8$.  } 
\end{figure}

\begin{figure}
  \begin{center}
\epsfxsize=10.cm
\begin{minipage}{\epsfxsize} \epsffile{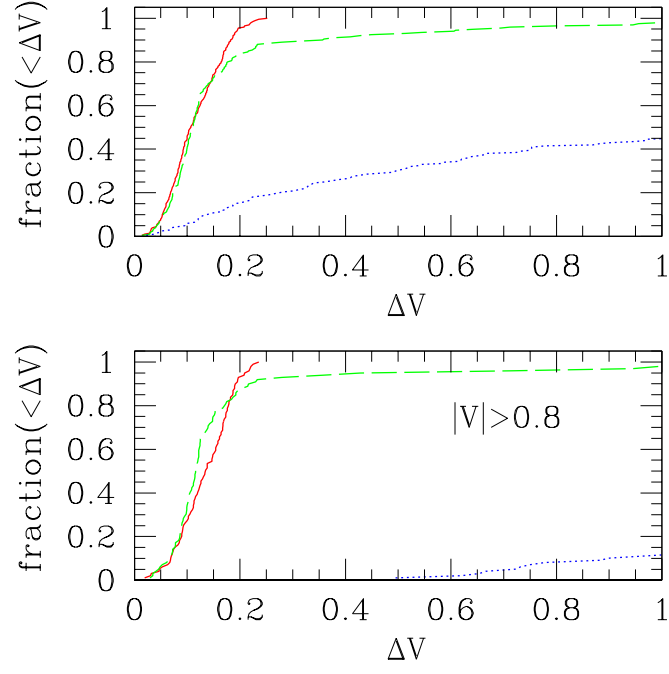} \end{minipage}
 \end{center}
  \caption{ Same as figure 4
 with LIGO-Hanford  and  VIRGO detectors (case (ii)). } 
\end{figure}

\begin{figure}
  \begin{center}
\epsfxsize=10.cm
\begin{minipage}{\epsfxsize} \epsffile{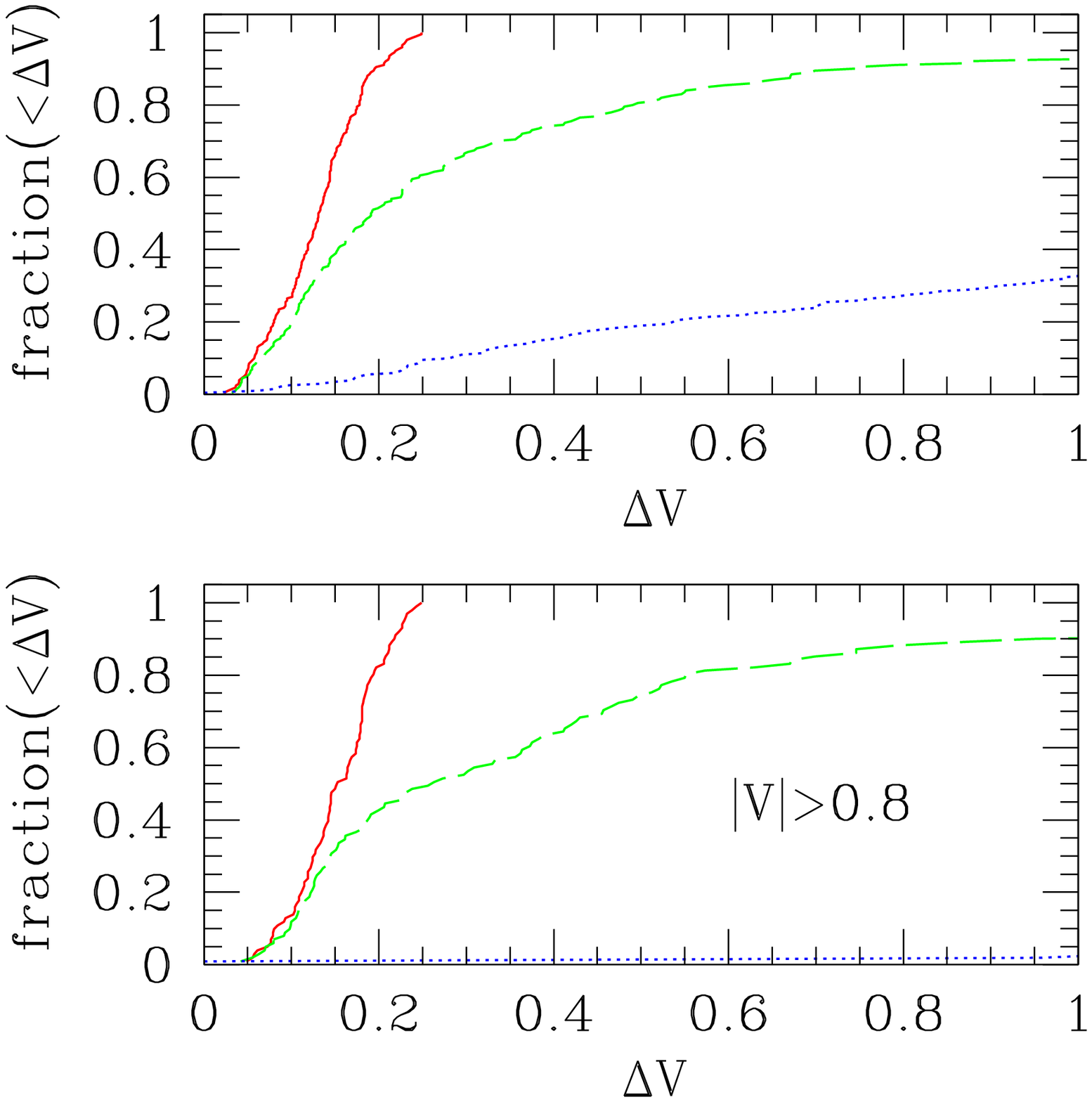} \end{minipage}
 \end{center}
  \caption{ Same as figure 4
 with two LIGO detectors (case (iii)). } 
\end{figure}

In this subsection we compare estimations errors $\Delta V$
given by $\lla \Delta V \Delta V \rra_N^{1/2}$, $\lla \Delta V \Delta V
\rra_r^{1/2}$  and $\lla \Delta V \Delta V\rra_C^{1/2}$ presented in the
previous subsection. The first one  $\lla \Delta V \Delta V \rra_N^{1/2}$ is
the prediction by Fisher matrix for a binary whose distance is not given
 from
EMW observation. The second one  $\lla \Delta V \Delta V
\rra_r^{1/2}$ is also predicted by Fisher matrix but for binaries with known
distances. For the third one we apply the $\cq$-$V$ constraint to
binaries with known distances. For each network (i)-(iii)  we prepare sample of  ``detectable binaries'' with $SNR_{tot}\ge 8$ in the
following manner.  From eq.(\ref{snr}) the maximum distance $r_{max}$ of
binaries with the threshold $SNR_{tot}=8$ is given as $r_{max}=r_0[2\max
\sigma(N)]^{1/2} /8$.  A binary with $SNR_{tot}\ge 8$ must be in a
sphere with radius $r_{max}$ around the 
Earth, since this distance $r_{max}$ is given for  optimal
configuration. In  this sphere we put a binary with random position and
orientation. If its 
$SNR_{tot}$ is larger than 8, we add it to our sample of ``detectable
binaries''. We continue this until the total number of our sample
becomes 200 for each network (i)-(iii). 

We find that, out of these samples, the numbers of binaries with
$|V|>0.8$ are 102 for case (i), 99 for case (ii) and 101 for case (iii).
For binaries with $|V|>0.9$ we have 64, 62 and 60 respectively.
While the intrinsic distribution of the parameter $|V|$ is homogeneous
in the range $[0,1]$ with random direction $\veN$ and orientation
${\vect L}$, our
``detectable 
binaries'' have skewed distributions of $|V|$ after selected with a signal-to-noise
threshold. 

In figure 5 we  show the distribution of the estimation errors $\Delta
V$ for case (i). The upper panel  shows the results for  all 200
binaries and the lower panel is only for 102 binaries with 
$|V|>0.8$. Similar results are given in figs. 6 and 7 for cases (ii) and
(iii). Comparing distributions for two Fisher matrix predictions $\lla
\Delta V \Delta V \rra_r^{1/2}$ (dashed curves) and $\lla \Delta V
\Delta V 
\rra_N^{1/2} $ (dotted curves), it is apparent that the estimation of
the parameter $V$ is significantly improved, if the distance $r$ to the
binary is determined by EMW observation.  The
differences between two curves become larger for subsample with
$|V|>0.8$ due to the degeneracy for the parameter fitting at the face on
limit $|V|\to 1$. 

We can study the effects of the $\cq$-$V$ constraint with solid curves and
dashed curves in figs.5-7. Note that this constraint does not always decrease the
estimation error $\Delta V$, as shown in figs. 5 and 6. However, the tail
of large $\Delta V$ are removed by the constraint. As a result, the
error $\Delta V$ becomes smaller than  $\sim 0.25$ for our samples. The
tails are mainly made by binaries with  $\epsilon \sim 1$ for which it
is difficult to observe two independent polarization modes and thereby
solve two parameters  $(V,\bpsi)$ separately. The constraint works well especially
for case 
(iii) with two LIGO detectors whose orientations are nearly aligned. In
this case the situation is similar to the observation with using a
single detector as studied  in \S 2.  In the lower panel of
figure 7 the median value of the error $\Delta V$ is reduced by $\sim
40\%$ by using the constraint.  This indicate that we should be careful
to use a simple 
 Fisher matrix approach for evaluating the parameter estimation error  for inclination.

\section{   Discussions}
We discussed  prospects of  a  method to constrain the inclination angle of a coalescing
 compact 
 binary  by detecting its gravitational waves associated with a three-dimensionally  localized SHB.
With this method we can get an important geometrical information to
understand the 
properties of SHBs and their 
afterglows as a function of the viewing angles of the jets (see {\it
e.g.} \cite{Yamazaki:2004ha} for long-soft bursts).
We should comment that there would be  a selection effect toward 
larger $|V|$ 
for a simultaneous detection of gravitational waves and SHB. This is
because the observed gravitational wave amplitude would be larger (see
\S IV.B) and
the SHB would be also luminous with the face-on configuration \cite{Nakar:2005bs,Kochanek:1993mw}.

 By analyzing
nearby samples with our method we might 
establish an efficient empirical criteria to select almost face-on
binaries using observed properties of  EMW signals.  Then  we can study
cosmological parameters with  face-on binaries at relatively
high redshift  where cosmological effects beyond the Hubble-law become
important.  The following is the outline of this approach. We first 
obtain the intrinsic 
amplitudes $A$ by putting  $R=R_{max}$ for the observed gravitational
wave amplitude $(AR)$ for binaries that are likely to be nearly face-on
according to the empirical criteria.
Then the
luminosity 
distances $r$ are obtained from eq.(10) and the estimated chirp masses.
 In  this manner  we
might observationally study the 
redshift-distance relation with the redshift information naturally obtained from
identified host galaxies with EMW observation. Note that this argument to estimate the
distance $r$ is different from the previous method in this paper to constrain the
inclination angle by simply converting the observed redshift to the distance $r$ for
nearby binaries.
Considering the strong correlation between the inclination $V$ and the
distance $r$ for parameter estimation \cite{Cutler:1994ys},  this  could
be a powerful approach to investigate the 
dark 
energy 
with  future gravitational wave detectors that can detect compact
binaries at 
$z\sim 1$ (see also \cite{Schutz:gp,Dalal:2006qt,Finn:1995ah}). 

\acknowledgments
The author would like to thank the referee for helpful comments to
improve the draft. He also thanks H. Takahashi for useful
conversations.  This research was funded by McCue Fund at the Center for Cosmology,  
UC Irvine.

\end{document}